\documentstyle[prl,aps,epsfig]{revtex}

\begin{document}
\twocolumn[
\hsize\textwidth\columnwidth\hsize\csname@twocolumnfalse\endcsname

\title{Ordering in a spin glass under applied magnetic field }
\author{Doroth\'ee Petit, L. Fruchter and I.A. Campbell }

\address{Laboratoire de Physique des Solides,\\
Universit\'e Paris Sud, 91405 Orsay,
France}

\maketitle

\begin{abstract}
Torque, torque relaxation, and magnetization measurements on a {\bf Au}Fe
spin glass sample are reported. The experiments carried out up to $7 T$ 
show a transverse
irreversibility line in the $(H,T)$ plane up to high applied fields, and a
distinct strong longitudinal irreversibility line at lower fields.  The data demonstrate for that this type of sample, a Heisenberg spin
glass with moderately strong anisotropy, the spin glass ordered state
survives under high applied fields in contrast to predictions of certain "droplet" type scaling models. The
overall phase diagram closely ressembles those of mean field or chiral 
models, which both have replica symmetry breaking transitions.

\end{abstract}

\twocolumn
\vskip.5pc ]
\narrowtext

For the infinite dimension or mean field spin glass, there is a true replica symmetry breaking (RSB) phase
transition under an applied field with, for Heisenberg spins, a transverse
irreversibility onset followed at lower fields by a crossover to
longitudinal irreversibility  \cite{AT,GT}. The mean field $(H,T)$ phase diagram including the
effect of anisotropy has been extensively studied theoretically. It has been strongly argued that
the physics of spin glasses below the upper critical dimension $d=6$ is basically similar to that in
infinite dimension \cite{parisi}. If real spin glasses in dimension three undergo an RSB transition, one should expect to find an in-field phase diagram qualitatively similar to that of mean field.
Alternatively if the standard Fisher-Huse scaling (or "droplet") scenario is physically correct, in three dimensions a true transition should exist only in zero
field \cite{fisher} and no irreversibilities should be seen under applied fields. The
experimental situation is not clear cut and is complicated by the fact that it has been hard to find a crucial physical measurement to rule out or alternatively to definitively establish the existence of an in-field frozen state. Magnetization experiments have been analysed in terms of transition lines in the $(H,T)$ plane\cite {monod,tholence,kenning}, while
susceptibility data (on an Ising like material) have been interpreted as
demonstrating an absence of ordering in a finite field \cite{mattsson}.

Torque measurements have the advantage of being directly sensitive to
transverse irreversibility. 
The Dzyaloshinski-Moriya (DM) interaction is the source of magnetic
anisotropy in spin glasses, leading to torque when an applied field is
turned \cite {fertlevy,hippert}. Because of the special character of the
spin glass anisotropy, this torque is observed only if there is a frozen-in
spin arrangement. If the spin glass is in a paramagnetic state, meaning that
the spins can reorganize themselves locally as soon as the field is turned,
there will be no torque for an isotropic polycrystalline sample. The torque
criterion for identifying a frozen spin glass state was exploited early on
over a restricted field range \cite{ian,albert,nadine}, but although the
question of spin glass ordering and aging has been addressed by
progressively more sophisticated magnetization and susceptibility
experiments (see \cite{jonason} for instance), the torque technique has been
neglected ; in particular there have been no systematic comparisons of
torque and magnetization on one and the same sample over an extended field
range. Thus the theoretical predictions have only been incompletely tested.
Here we report extensive torque, torque relaxation, and magnetization
measurements to high fields on a sample of the archetype spin glass,
{\bf Au}Fe. The torque data show a clear transverse irreversibility
transition line below which the spin arrangement remains frozen over very
long times even under strong applied fields. The magnetization data indicate
a quite distinct strong longitudinal irreversibility line. The experimental in-field
phase diagram bears a striking qualitative resemblance to that of the
Heisenberg mean field model with strong anisotropy\cite{GT,kotliar}.

We studied a sample of {\bf Au}5{\%} Fe prepared by standard metallurgical
techniques. {\bf Au}Fe is a Heisenberg spin glass with moderately strong DM anisotropy
\cite{alloul}. The sample was heavily cold worked and then annealed to
guarantee homogenity. The $T_g$ estimated with applied field extrapolated to zero is
 $20.6 K$. The torque measurements were performed using a capacity method ; applied fields
up to $7 T$ were provided by a horizontal superconducting Helmhotz coil. The
main experimental difficulty was eliminating parasitic signals arising from
the interaction of the sample moment with a residual field gradient from the
coils. Magnetization measurements were carried out on a commercial SQUID
instrument.

\begin{figure}
\centerline{\includegraphics*[width=1\hsize]{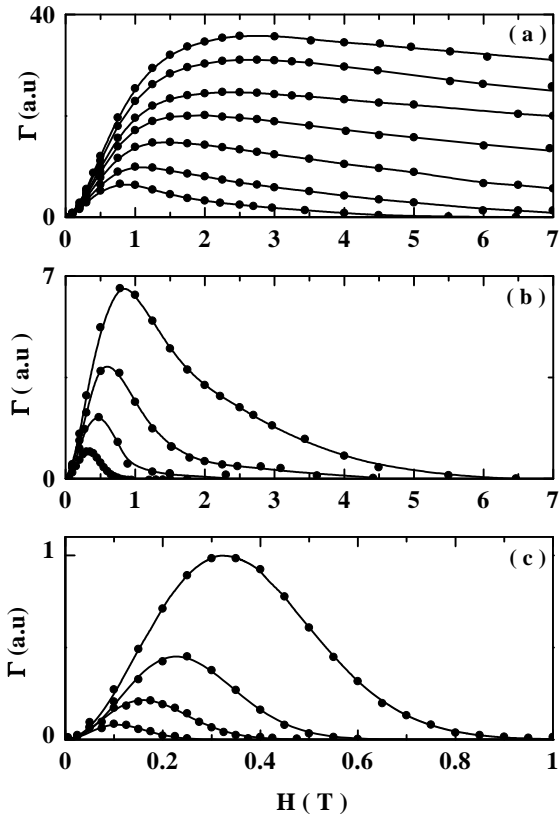}}
\caption{The torque measured 30 seconds after field cooling followed by 
turning the field by $5^\circ$. Each point correspond to a separate field 
cooling run. The torque is in arbitrary units but the units are the same 
for the three plots. The temperatures are, from top to bottom: 
(a) $5 K$, $6 K$, $7 K$, $8 K$, $9 K$, $10 K$, $11 K$ ; (b) $11 K$, $12 K$, $13 K$, $14 K$ ; 
(c) $14 K$, $15 K$, $16 K$, $17 K$}
\label{couplechamp}
\end{figure}

The principal protocol used for the torque measurements was to field cool
(FC) the sample in an applied field $H$ to the measuring temperature $T$ ;
once the temperature was established, the field was turned, typically by an
angle of $5^\circ$.The torque was measured from a few seconds after the turn
and for times up to an hour. We will first describe the overall pattern of
the torque signals as a function of $H$ and $T$. Fig.\ref{couplechamp} shows the observed
torque values ; each point corresponds to a separate FC run.We have chosen
to plot the points measured at 30 seconds after the field was turned.
Relaxation effects will be discussed later on. 

The DM anisotropy is due to a sum of terms of the form ${\bf D}_{ij}({\bf
S}_i\wedge{\bf S}_j)$
\cite{fertlevy}. Each time a sample is cooled either in field or in zero
field, the spins conspire to minimize the total spin-spin interaction plus
anisotropy energy by taking up an appropriate configuration. Once a rigid
configuration has been established, turning it bodily costs energy leading
to anisotropy with respect to its original orientation. If the spins can
completely rearrange, they can take up a configuration which is different on
the microscopic level, so the anisotropy reorients, and the torque disappears.
Zero torque thus indicates a paramagnetic state\cite{ian}.

Suppose that a spin glass has a strictly rigid spin configuration with
magnetization $M(H)$ and a field independent spin glass anisotropy $K$. Then
the torque signal $\Gamma$ when the applied field $H$ is turned by an angle
$\theta$ is given by\cite{hippert}
\begin{equation}
\theta/\Gamma=1/K +1/M_rH
\end{equation}
For a series of points each taken after cooling in field, the torque signal
$\Gamma(H)$ will initially increase with field as $H^2$ because the
magnetization is proportional to $H$; when the limit $HM >> K$ is reached
the torque will saturate at a field independent value depending only on
$Ksin\theta$ \cite{hippert}. This is what is observed at the lowest
temperatures in Fig.\ref{couplechamp}. However at higher temperatures, the observed
torque signal initially increases with field as at low temperatures ; it
reaches a peak at a field $H_p(T)$ and then for higher fields it decreases
again until it becomes unobservably small at a critical field $H_c(T)$.

Thus the data show that at low temperatures, $K$ tends to become field
independent for the range of fields available to us, i.e. the spin
configuration is almost rigid and the DM anisotropy after cooling in field is
almost independent of the value of the field. With increasing temperature
the low field behaviour is still of the same form so $K(T)$ is still essentially field
independent, but $K(T)$ decreases regularly. This is because local
spin-wave-like fluctuations reduce the time average effective local spin
moments $<{\bf S}_i>$, so each term in the DM expression above becomes
weaker on increasing temperature. Higher fields lead to the peak effect,
showing that a combination of temperature and field begins to weaken the
rigid state. The spins are still frozen but they have become free to select
configurations for which the DM terms are weaker, producing a progressive
reduction in $K(H,T)$ with field. Finally above $H_c(T)$ the system can
completely rearrange the spins on a time short compared with the time scale
of the measurement, and there is no more observable anisotropy. Above this
critical field the system has entered the paramagnetic state.( The precise
form of the behaviour of a spin glass with weak anisotropy may well be
rather different). This measurement gives detailed information
on the progressive manner by which a spin glass system loses rigidity under
increasing applied fields and temperatures.

Individual points on the $H_c(T)$ curve were estimated by plotting $log(\Gamma(H,T))$ against $T$ and observing the intersection with the noise level, which was typically $0.01$ in the units of Figure 1. We have also estimated the position of the critical line $H_c(T)$ in a
complementary and more sensitive way by measuring the torque signal as a
function of time after turning. $H_c(T)$ is then defined as the field above which 
there is no observable torque relaxation (so no observable torque above the noise). 
The two sets of estimates are entirely consistent. Error bars are indicated on Figure 2.

Magnetic measurements provide an alternative method for identifying an
irreversibility line. Field cooled and zero field cooled (ZFC)
magnetizations are compared ; the onset of difference between the two
indicates irreversibility\cite{monod,tholence,kenning}. For a {\bf Cu}Mn
sample Kenning, Chu and Orbach\cite{kenning}
observed " strong " and " weak " irreversibility lines ; they identified the
latter with a transverse irreversibility as had been seen in torque measurements
over a restricted range of fields \cite{nadine}. We have carried out
magnetization measurements on the present sample. Following Kenning
et al, we have plotted the difference between $M_{FC}$ and $M_{ZFC}$. $[M_{FC}-M_{ZFC}]/M_{FC}(5K)< 10^{-3}$ gives a criterion which defines an effective critical temperature at each field. (In any case, theory suggests this line is a crossover and so intrinsically fuzzy).To the precision of our SQUID measurements, we
could not observe a weak irreversibility line, and our critical points
$H_{cm}(T)$ correspond to the strong irreversibility of Kenning et al. It
can be noted that while the torque gives a transverse irreversibility
criterion for $H_c(T)$ which is very clear cut experimentally, the weak
longitudinal irreversibilty criterion of Kenning et al requires painstaking
measurements of tiny magnetization differences between successive FC and ZFC
runs. Even the strong irreversibility signal becomes small at high fields
(less than 1 percent of $M_{FC}$ at $6K$ by $3T$). Although the transverse
irreversibility line can be taken as representing a true transition, it is a "stealthy " transition - essentially invisible in any longitudinal measurement, whether by
magnetization differences or a.c. susceptibility. This implies that except
in low fields, no longitudinal magnetization measurement can be used as a
reliable probe of the onset of true spin freezing, and transverse
irreversibility must be studied in order to establish an $(H,T)$ phase diagram.

\begin{figure}[htb]
\centerline{\includegraphics*[width=1\hsize]{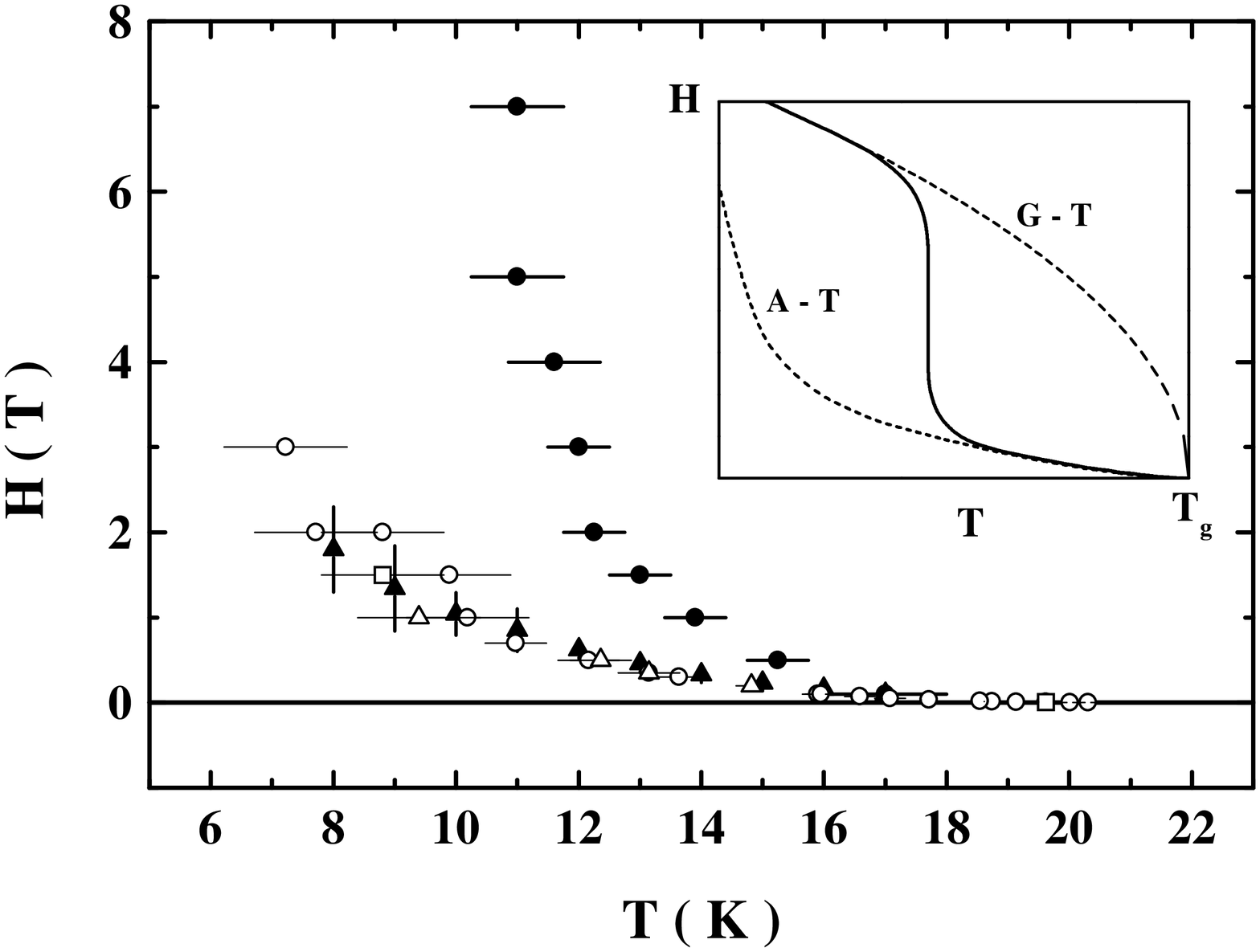}}
\caption{Irreversibility onsets estimated for transverse irreversibility 
($H_c(T)$ from torque measurements, full circles) and longitudinal irreversibility 
($H_{cm}(T)$ from SQUID measurements, open circles, squares and triangles). Open circles 
correspond to temperature increment of $0.2 K$, squares to $0.1 K$, triangles to $0.25 K$. 
The full triangles correspond 
to the torque peaks $H_p(T)$ of Fig.\ref{couplechamp}. The inset is a 
schematic drawing of the mean field phase diagram with longitudinal 
d'Almeida-Thouless (AT) [1] and transverse Gabay-Toulouse (GT) [2]  irreversibility 
onset lines. For a sample with strong random anisotropy, theory predicts a 
transverse irreversibility onset with a crossover, following the full line 
(Kotliar and Sompolinski [15])}
\label{diagramme}
\end{figure}

The results for the phase diagram using these alternative criteria are
displayed in Fig.\ref{diagramme}. The $H_c(T)$ line is of similar form to that already
observed in torque measurements on a {\bf Au}2{\%}Fe sample at  low
fields \cite{nadine}; the present results extend the torque data by an order of magnitude in field range and provide longitudinal measurements on one and the same sample. The $H_c(T)$ form is characteristic of RSB predictions for samples with strong
anisotropy, where the transverse transition line follows AT behaviour at low fields and then crosses over towards GT like behaviour at high fields \cite{kotliar}, see inset. Clearly the present $H_c(T)$ line ressembles the full line in the inset, reaching the crossover region but not the GT limit.

We can note that for this sample the peak field $H_p(T)$ line from the
torque experiments lies very close to the longitudinal irreversibility line.

Although qualitative agreement between theory and experiment appears excellent there is an important caveat. It would not appear meaningful to use the standard model to analyse the field dependences of the transition temperature, because the three dimension Heisenberg spin glass transition
calculated with the standard Edwards Anderson order parameter is already at zero temperature in zero field \cite{olive,kawamura1}, so an alternative model must be sought.

A most attractive explanation for the observation of the finite
temperature transitions in real Heisenberg spin glasses is that of Kawamura
\cite{kawamura1} who proposes that the transition is fundamentally chiral,
and that it is " revealed " by the presence of even weak anisotropy. Calculations show that the chiral model $(H,T)$ transition behaviour is of an RSB type, and mimics the mean
field behaviour \cite{kawamura2,kawamura9}. For fields strong compared to the
anisotropy the transverse irreversibility transition line lies at $H_c
\propto (T-T_g)^{0.5}$ \cite{kawamura2} as for the GT line. For low fields
$H_c \propto (T-T_g)^\phi$ with $\phi$ between 1.3 and 1.5 
much as for the AT line \cite{kawamura9}. There is an anisotropy dependent crossover from AT-like behaviour to GT-like as in the mean field model. In the chiral model the transition irreversibility line is a true transition line, but the transition may well be of a very different nature from that in the mean field model (it might be one step RSB for instance \cite{kawamura2}).
The present transverse irreversibility data are completely compatible with the chiral
model predictions for the irreversibility onset if we consider that the
steep rise at lower temperatures indicates that even at $7 T$ the system is
still in a crossover regime and the true GT-like regime would require yet
stronger fields (c.f.\cite{kotliar}). The longitudinal irreversibility
$H_{cm}$ line follows an AT like behaviour with exponent $\phi$ about 1.5
from $T_g$ to near $0.8 T_g$ and then takes a larger exponent.

\begin{figure}
\centerline{\includegraphics*[width=1\hsize]{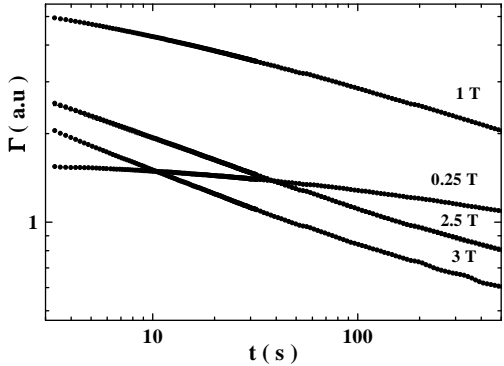}}
\caption{Torque relaxation as a function of time in seconds after field 
cooling followed by turning by $5^\circ$ at temperature 11 K. The values 
of the applied field are indicated on the figure. A straight line on a log-log plot 
corresponds to algebraic decay of the torque signal.}
\label{relax}
\end{figure}

We now turn briefly to the question of relaxation. In the region below
$H_c(T)$ the torque signal always relaxes
with time in the algebraic or quasi-logarithmic manner familiar from spin
glass magnetization
measurements, Fig.\ref{relax}. This is true above and below the $H_{cm}(T)$ line. This
form of signal decay means that
there is no maximum characteristic time for the relaxation, which is a
criterion indicating that the system is in a glassy 
frozen state and that the signal decay reflects a form of magnetic creep. We
conclude that when the torque signal is observable, the system is frozen in
this sense ; there is a line of true freezing transitions at or very near 
to the $H_c(T)$ line in Fig.\ref{diagramme}. Though chiral model simulations have so far only been carried out in zero field, the slow quasi-algebraic form of relaxation found experimentally appears compatible with the equilibrium simulation relaxation data \cite{kawamura2}.

The experimental torque aging effects after cooling in field appear to be
negligible (as in {\bf Cu}Mn \cite{carlo}), in contrast to those always
observed in spin glass magnetization experiments at zero field 
\cite{vincent}, and to the aging observed in the zero field chiral simulations \cite{kawamura2}. Simulations to check for an in-field supression of aging in the chiral approach would provide an important verification of the model.

In conclusion, by combining torque and magnetization information over a wide
range of applied fields, we find that a Heisenberg spin glass with strong
random DM anisotropy has an in-field phase diagram which is remarkably
similar to that of the chiral ordering model \cite{kawamura1}(which mimics the well established mean field 
type \cite{GT,kotliar}): a transverse irreversibility onset line
corresponding to a
true RSB freezing transition, plus a lower and quite distinct strong longitudinal
irreversibility line which can be identified from magnetization 
data. The two lines fuse at low applied fields. 
To make a realistic quantitative comparison with theory concerning characteristics like the in-field aging and the transverse magnetization decay, we must await full in-field simulations in the chiral ordering scenario \cite{kawamura2}
with strong anisotropy. Already the striking qualitative similarities
between the experimental phase diagram and the chiral (or mean field) spin
glass model phase diagram is strong evidence that the essential physics of
real life Heisenberg spin glasses is very close to that of the RSB class of models. Scaling approaches of the Fisher-Huse droplet type \cite{fisher} do not appear to be 
compatible with the experimental data, as they
predict that whenever the applied field is non-zero, there will be 
no transition to a frozen state at any finite temperature.

Acknowledgements. We would like to thank Professor H. Kawamura for very
helpful information on the chiral model.


\begin{references}



\bibitem{AT} J.R.L. d'Almeida and D.J. Thouless, J. Phys. {\bf A 11}, 983
(1978),
\bibitem {GT} M. Gabay and G. Toulouse, Phys. Rev. Lett. {\bf 47}, 201 
(1981)
\bibitem{parisi} E. Marinari, G. Parisi and J.J. Ruiz-Lorenzo, " Spin
Glasses and Random fields ", {\it Series on Directions in Condensed Matter
Physics} Vol 12, A.P. Young (Editor), World Scientific 1998, p 59
\bibitem{fisher} D.S. Fisher and D.A. Huse, Phys. Rev. B {\bf 38}, 373
(1988)
\bibitem{monod} P. Monod and H. Bouchiat, J. Phys (Paris) Lett. {\bf 43},
L145 (1982)
\bibitem{tholence} J.L. Tholence and M. B. Salomon, J. Mag. Mag. Mat. 
{\bf 31-34}, 1340 (1983)
\bibitem{kenning} G.G. Kenning, D. Chu and R. Orbach, Phys. Rev. Lett. 
{\bf 66}, 2923 (1991)
\bibitem{mattsson}J. Mattsson, T. Jonsson, P. Nordblad, H. Aruga Katori and
A. Ito, Phys. Rev. Lett. {\bf 74}, 4305 (1995)
\bibitem{fertlevy} A.Fert and P.M. Levy, Phys. Rev. Lett. {\bf 44}, 1538
(1980)
\bibitem{hippert} A. Fert and F. Hippert, Phys. Rev. Lett. {\bf 49}, 1508
(1982)
\bibitem{ian} I.A. Campbell, N. de Courtenay and A. Fert, J. Phys. Lettres
{\bf45}, L565 (1984)
\bibitem{albert} A. Fert, N. de Courtenay and I.A. Campbell, J. Appl. Phys.
{\bf 57}, 398 (1985)
\bibitem{nadine} N. de Courtenay, A. Fert and I.A. Campbell, Phys. Rev.B
{\bf 30}, 6791 (1984)
\bibitem{jonason} K. Jonason, E. Vincent, J. Hammann, J.P. Bouchaud, and P.
Nordblad, Phys. Rev. Lett. {\bf 81}, 3243 (1998)
\bibitem{kotliar} G. Kotliar and H. Sompolinsky, Phys. Rev. Let. {\bf53},
1751 (1984)
\bibitem{alloul} H. Alloul and F. Hippert, J. Phys. {\bf 41}, L201 (1980)
\bibitem{olive} J.A. Olive, A. P. Young and D. Sherrington, Phys. Rev. B
{\bf 34}, 6341 (1986)
\bibitem{kawamura1} H. Kawamura, Phys. Rev. Lett. {\bf 68}, 3785 (1992)
\bibitem{kawamura2} H. Kawamura, Phys. Rev. Lett. {\bf 80}, 5421 (1998), H.Kawamura and K. Hukushima, "Computer Simulation Studies 
in Condensed-Matter Physics XI", Springer,Berlin (1999) 7, cond-mat/9805118, and unpublished 
\bibitem{kawamura9} H. Kawamura, unpublished.
\bibitem{carlo} C. Giovanella, I.A. Campbell, and F. Hippert, Phys. Rev.
Lett. {\bf 59}, 1030 (1987)
\bibitem{vincent} M. Lederman, R. Orbach, J. Hammann, M. Ocio and E. Vincent,
Phys. Rev. B {/bf 44}, 7403 (1991)

\end{references}
\end{document}